\documentclass[12pt,aps,prd,preprint,onecolumn]{revtex4}
\usepackage{amssymb}
\usepackage{amsmath}
\usepackage{latexsym}
\usepackage{array}
\usepackage[activeacute,english]{babel}
\usepackage[latin1]{inputenc}
\usepackage[dvips]{graphicx}
\usepackage{color}
\usepackage{epsfig}

\newcommand{\bea}{\begin{eqnarray}}
\newcommand{\eea}{\end{eqnarray}}
\newcommand{\be}{\begin{equation}}
\newcommand{\ee}{\end{equation}}

\newcommand{\bc}{\begin{center}}
\newcommand{\ec}{\end{center}}
\newcommand{\la}{\label}

\newcommand{\nn}{\nonumber}
\newcommand{\bi}{\bibitem}

\begin{document}

\title{\Large \bf Particle mass generation from physical vacuum}

\author{Carlos Quimbay\footnote{Associate researcher of Centro
Internacional de F\'{\i}sica, Bogot\'a D.C., Colombia.}}
\email{cjquimbayh@unal.edu.co}

\affiliation{Departamento de F\'{\i}sica, Universidad Nacional de Colombia.\\
Ciudad Universitaria, Bogot\'{a} D.C., Colombia.}

\author{John Morales\footnote{Associate researcher of Centro
Internacional de F\'{\i}sica, Bogot\'a D.C., Colombia.}}
\email{jmoralesa@unal.edu.co}

\affiliation{Departamento de F\'{\i}sica, Universidad Nacional de Colombia.\\
Ciudad Universitaria, Bogot\'{a} D.C., Colombia.}

\date{\today}

\begin{abstract}
We present an approach for particle mass generation in which the
physical vacuum is assumed as a medium at zero temperature and
where the dynamics of the vacuum is described by the Standard
Model without the Higgs sector. In this approach fermions acquire
masses from interactions with vacuum and gauge bosons from charge
fluctuations of vacuum. The obtained results are consistent with
the physical mass spectrum, in such a manner that left-handed
neutrinos are massive. Masses of electroweak gauge bosons are
properly predicted in terms of experimental fermion masses and
running coupling constants of strong, electromagnetic and weak
interactions. An existing empirical relation between the top quark
mass and the electroweak gauge boson masses is explained by means
of this
approach. \\

\vspace{0.4cm} \noindent {\it{Keywords:}} {Particle mass
generation, physical vacuum, Standard Model without Higgs sector,
self-energy, polarization tensor.}

\end{abstract}

\maketitle


\section{Introduction}\label{sec:01}

The Higgs mechanism is the current accepted procedure to generate
masses of electrically charged fermions and electroweak bosons in
particle physics \cite{djouadi}. The implementation of this
mechanism requires the existence of a sector of scalar fields
which includes a Higgs potential and Yukawa terms in the
Lagrangian density of the model. In the Minimal Standard Model
(MSM), the Higgs field is a doublet in the $SU(2)_L$ space
carrying a non-zero hypercharge, and a singlet in the $SU(3)_C$
space of color. The Higgs mechanism is based on the fact that the
neutral component of the Higgs field doublet spontaneously
acquires a non-vanishing vacuum expectation value. Since the
vacuum expectation value of Higgs field is different from zero,
the Higgs field vacuum can be interpreted as a medium with a net
weak charge. In this way the $SU(3)_C \times SU(2)_L \times
U(1)_Y$ symmetry is spontaneously broken into the $SU(3)_C \times
U(1)_{em}$ symmetry \cite{weinberg}.

As a consequence of the MSM Higgs mechanism, the electroweak gauge
bosons acquire their masses so that the masses depend on the
vacuum expectation value of Higgs field, which is a free parameter
in the MSM. This parameter can be fixed by means of calculating
the muon decay at tree level using the Fermi effective coupling
constant. Simultaneously, Yukawa couplings between the Higgs field
and fermion fields lead to the generation of masses for
electrically charged fermions that depend on Yukawa coupling
constants, which also are free parameters in the MSM. These
constants can be fixed by means of experimental values of
fermionic masses. The above mechanism implies the existence of a
neutral Higgs boson in the physical spectrum being its mass a free
parameter in the MSM. Because it just exists left-handed neutrinos
in the Lagrangian density of the MSM, then neutrinos remain
massless after the spontaneously electroweak symmetry breaking.

In the current picture of Higgs mechanism \cite{djouadi}, masses
of the MSM particles spectrum are generated through interactions
between electroweak gauge bosons and electrically charged fermions
with weakly charged Higgs field vacuum. However there are some
physical aspects in this picture of mass generation that are not
completely satisfactory summarized in the following questions:
What is the possible description of interactions between fermions
and electroweak bosons with Higgs field vacuum? How possible is it
to show fundamentally that particle masses are generated by these
interactions? Why is it that the origin of particle masses is just
related to the weak interaction? Why is it that the most-intense
interactions (strong and electromagnetic) are not related to the
mass generation mechanism? Why are there no interactions between
weakly charged left-handed neutrinos and the weakly charged Higgs
field vacuum? Why left-handed neutrinos are massless if they have
a weak charge?

All the above questions might have a trivial answer if we only
look at things through the current picture of Higgs mechanism.
However we are interested in exploring a possible physics behind
the Higgs mechanism. On this manner we propose an approach for
particle mass generation in which fermions acquire their masses
from theirs interactions with physical vacuum and gauge bosons
from charge fluctuations of vacuum.

Physical vacuum is the state of lowest energy of all gauge bosons
and fermion fields \cite{schwinger1}. As it is well known from the
covariant formulation of Quantum Field Theory \cite{schwinger2},
the state of lowest energy of gauge bosons and matter fields has
an infinite energy. This physical vacuum is a rich medium where
there are processes involving particles and antiparticles with
unlimited energy. The physical vacuum is then assumed as a medium
at zero temperature which is formed by fermions and antifermions
interacting among themselves by exchanging gauge bosons. From a
cosmological point of view the physical vacuum can be assumed as
an almost equilibrated medium which corresponds to an infinitely
evolving vacuum \cite{bjorken,volovik}.

The fundamental model describing the dynamics of physical vacuum
is the Standard Model without the Higgs Sector (SMWHS), which is
based on the $SU(3)_C \times SU(2)_L \times U(1)_Y$ gauge symmetry
group. We assume that each fermion in the physical vacuum has
associated a chemical potential which describes the excess of
antifermions over fermions. Then there are twelve fermionic
chemical potentials $\mu_{f}$ associated with the six leptons and
the six quarks implying an antimatter-matter asymmetry in the
physical vacuum. Hence the physical vacuum is considered as a
virtual medium having antimatter finite density. This
antimatter-matter asymmetry of physical vacuum is related to CP
violation by electroweak interactions. Naturally the mentioned
asymmetry has an inverse sign respect to the one of the
matter-antimatter asymmetry of Universe. The existence of
fermionic chemical potentials in physical vacuum does not imply
that this vacuum itself carries net charges. This can be
understood in a similar way as the existence of the maximal
matter-antimater asymmetry of Universe which does not mean that
the baryonic matter carries net charges. This idea to have
fermionic vacuum densities responsible for distinct fermion masses
has also been suggested by \cite{bopp}.

The masses of fermions are obtained starting from their
self-energies which give account of the fundamental interactions
of massless fermions with physical vacuum \cite{schwinger1}. While
quark masses are generated by strong, electromagnetic and weak
interactions, the electrically charged lepton masses are only
generated by electromagnetic and weak interactions and the
neutrino masses are generated from the weak interaction. On the
other hand, gauge boson masses are obtained from charge
fluctuations of the physical vacuum which are described by the
vacuum polarization tensors \cite{schwinger1}.

We use the following general procedure to calculate particle
masses: Initially we write one-loop self-energies and one-loop
polarization tensors at finite temperature and density, next we
calculate dispersion relations by obtaining the poles of fermion
and gauge boson propagators, from these dispersion relations we
find fermion and gauge boson effective masses at finite
temperature and density, finally we identify these effective
masses at zero temperature with physical masses. This
identification can be performed because the virtual medium at zero
temperature is representing the physical vacuum.

From a different perspective other works have intended to show
that the inertial reaction force appearing when a macroscopic body
is accelerated by an external agent \cite{arueda1}. This reaction
force is originated as a reaction by the physical vacuum that
opposes the accelerating action \cite{arueda1}. All these works
involve just the electromagnetic quantum vacuum and have been able
to yield the expression $F=ma$ as well as its relativistic
generalization. An expression for the contribution by the
electromagnetic quantum vacuum to the inertial mass of a
macroscopic object has been found and this has been extended to
the gravitational case. Originally these works have used a
semiclassical approach \cite{arueda1} which has been easily
extended to a quantum version \cite{arueda2}.

We find that the fermion and gauge boson masses are functions of
the vacuum fermionic chemical potentials $\mu_{f}$ which are fixed
using experimental fermion masses. From the values of all
fermionic chemical potentials obtained we calculate the masses of
electroweak gauge bosons obtaining an agreement with their
experimental values. In this approach for particle mass generation
is obtained that left-handed neutrinos are massive because they
have a weak charge. The weak interaction among massless neutrinos
and the physical vacuum is the source of neutrino masses.
Additionally this approach can explain an existing empirical
relation between the top quark mass and the electroweak gauge
boson masses.

Before considering the case of physical vacuum described by the
SMWHS, in section II we first show how to obtain gauge invariant
masses of fermions and gauge bosons for the case in which the
dynamics of the vacuum is described by a non-abelian gauge theory.
In section III we consider the SMWHS as the model which describes
the dynamics of physical vacuum and we obtain fermion (quark and
lepton) and electroweak gauge boson ($W^{\pm}$ and $Z^0$) masses.
We have consistently the masses of electroweak gauge bosons in
terms of the masses of fermions and running coupling constants of
the three fundamental interactions. In section IV we focus our
interest in finding a restriction about the possible number of
families. Additionally we predict the mass of the quark top and a
highest value for the sum of the square of neutrino masses. Our
conclusions are summarized in section V.


\section{Non-abelian gauge theory case}\label{sec:02}

In this section we first consider a more simple case in which the
dynamics of vacuum is described by a non-abelian gauge theory, and
in this context we calculate fermion and gauge boson masses. The
vacuum is assumed to be a quantum medium at zero temperature
constituted by fermions and antifermions interacting among
themselves through non-abelian gauge bosons. We also assume that
there exist an excess of antifermions over fermions in vacuum.
This antimatter-matter asymmetry of vacuum is described by
non-vanishing fermionic chemical potentials $\mu_{f_i}$, where
$f_i$ represents different fermion species. In this section, for
simplicity we will take $\mu_{f_1}= \mu_{f_2}= \ldots = \mu_f$.

The non-abelian gauge theory describing the dynamics of vacuum is
given by the following Lagrangian density \cite{weldon}

\be {\cal L}=-\frac{1}{4} F_{A}^{\mu \nu} F_{\mu \nu}^{A} +
\bar{\psi}_m \gamma^{\mu} \left( \delta_{mn} i \partial_{\mu} + g
L_{mn}^{A} A_{\mu}^{A} \right) \psi_n , \la{lag} \ee where $A$
runs over the generators of the gauge group and $m,n$ run over the
states of the fermion representation. The covariant derivative is
$\mbox{D}_\mu = \delta_\mu + igT_A A_{\mu}^A$, being $T_A$ the
generators of the $SU(N)$ gauge group and $g$ the gauge coupling
constant. The representation matrices $L_{mn}^{A}$ are normalized
by $Tr(L^A L^B) = T(R) \delta^{AB}$ where $T(R)$ is the index of
the representation. In the calculation of fermionic self-energy
appears $(L^A L^A)_{mn} = C(R) \delta_{mn}$, where $C(R)$ is the
quadratic Casimir invariant of the representation \cite{weldon}.

At finite temperature and density, Feynman rules for vertices are
the same as those at $T=0$ and $\mu_f=0$, while propagators in the
Feynman gauge for massless gauge bosons $D_{\mu \nu}(p)$, massless
scalars $D(p)$ and massless fermions $S(p)$ are \cite{kobes} \bea
D_{\mu \nu}(p) &=& -g_{\mu \nu} \left[ \frac{1}{p^2+i\epsilon} -i
\Gamma_b(p) \right],  \la{bp} \\
D(p) &=& \frac{1}{p^2+i\epsilon}-i{\Gamma}_b(p),  \la{ep} \\
S(p) &=& \frac{p{\hspace{-1.9mm}\slash}}{p^2+i \epsilon}+ i
p{\hspace{-1.9mm} \slash}{\Gamma}_f(p),  \la{fp} \eea where $p$ is
the particle four-momentum and the medium temperature $T$ is
introduced through the functions $\Gamma_b(p)$ and $\Gamma_f(p)$
which are given by \bea
\Gamma_b (p)= 2\pi \delta(p^2)n_b (p),  \la{db} \\
\Gamma_f (p)= 2\pi \delta(p^2)n_f (p),  \la{df} \eea with \bea
n_b (p) &=& \frac{1}{e^{|p\cdot u|/T}-1}, \la{nb}\\
n_f(p) &=& \theta(p\cdot u)n_{f}^{-}(p)+\theta(-p\cdot
u)n_{f}^{+}(p), \la{nf} \eea  being $n_b(p)$ the Bose-Einstein
distribution function. Fermi-Dirac distribution functions for
fermions $n_{f}^{-}(p)$ and for anti-fermions $n_{f}^{+}(p)$ are
\bea n_{f}^{\mp}(p)= \frac{1}{e^{(p\cdot u \mp \mu_f)/T}+1}. \eea
In the distribution functions $(\ref{nb})$ and $(\ref{nf})$,
$u^{\alpha}$ is the four-velocity of the center-of-mass frame of
the medium, with $u^\alpha u_\alpha =1$.


\subsection{Self-energy and fermion mass}

We first consider the propagation of a massless fermion in a
medium at finite temperature and density. The finite density of
the medium is associated with the fact that medium has more
antifermions than fermions. The fermion mass is calculated by
following the general procedure that we have described in the
introduction.

For a non-abelian gauge theory with parity and chirality
conservation, the real part of the self-energy for a massless
fermion is written as \be \mbox{Re}\,\Sigma^{\prime}(K)=-
aK{\hspace{-3.1mm}\slash}-b u{\hspace{-2.1mm} \slash},  \la{tse}
\ee $a$ and $b$ are Lorentz-invariant functions and $K^{\alpha}$
the fermion momentum. These functions depend on Lorentz scalars
$\omega$ and $k$ defined as {\hspace {0.1 cm}}
$\omega\equiv(K\cdot u)$ and $k\equiv[(K\cdot u)^2-K^2]^{1/2}$.
For convenience $u^\alpha=(1,0,0,0)$ and then we have $K^2
=\omega^2-k^2$, where $\omega$ and $k$ can be interpreted as the
energy and three-momentum respectively. From $(\ref{tse})$ we can
write \bea a(\omega,k) &=& \frac{1}{4k^2} \left[
Tr(K{\hspace{-3.1mm}\slash}\, \mbox{Re}\,\Sigma^{\prime})- \omega
Tr(u{\hspace{-2.1mm}
\slash}\,\mbox{Re}\,\Sigma^{\prime}) \right],  \la{lifa} \\
b(\omega,k) &=& \frac{1}{4k^2} \left[ (\omega^2-k^2)
Tr(u{\hspace{-2.1mm}\slash}\,\mbox{Re}\,\Sigma^{\prime})- \omega
Tr(K{\hspace{-3.1mm}\slash}\,\mbox{Re}\,\Sigma^{\prime}) \right].
\la{lifb} \eea

The fermion propagator including only mass corrections is given by
\cite{weldon1} \be S(p)=\frac
1{K{\hspace{-3.1mm}\slash}-\mbox{Re}\,\Sigma ^{\prime }(K)}=
\frac{1}{r}\frac{\gamma^0 \omega n - \gamma_i k^i}{n^2 \omega^2 -
k^2}, \la{pft} \ee where $n = 1 + b(\omega,k)/r\omega$ and $ r = 1
+ a(\omega,k)$. Propagator poles can be found when \be \left[ 1 +
\frac{b(w,k)}{w(1 + a(w,k))} \right]^2 w^2 - k^2 = 0.  \la{fdr0}
\ee We observe in $(\ref{fdr0})$ that $n$ plays a role similar to
that of the index of refraction in optics. To solve the equation
$(\ref{fdr0})$, $a(\omega,k)$ and $b(\omega,k)$ are first
calculated from the relations $(\ref{lifa})$ and$(\ref{lifb})$ in
terms of the real part of the fermionic self-energy. The
contribution to the fermionic self-energy from the one-loop
diagram which can be constructed in this theory is given by \be
\Sigma (K)=ig^2C(R)\int \frac{d^4p}{(2\pi )^4}D_{\mu \nu }(p)
{\gamma }^\mu S(p+K){\gamma }^\nu ,  \la{fse} \ee where $g$ is the
interaction coupling constant and $C(R)$ is the quadratic Casimir
invariant of the representation. For the fundamental
representation of SU(N), $C(R)=(N^2-1)/2N$ \cite{weldon11}. We
have that $C(R)=1$ for the U(1) gauge symmetry group, $C(R)=1/4$
for SU(2) and $C(R)=4/3$ for SU(3).

Substituting $(\ref{bp})$ and $(\ref{fp})$ into $(\ref{fse})$, the
fermionic self-energy can be written as
$\Sigma(K)=\Sigma(0)+\Sigma^{\prime}(K)$, where $\Sigma(0)$ is the
zero-density and zero-temperature contribution and
$\Sigma^{\prime}(K)$ is the contribution at finite temperature and
density. Then we have that \bea \Sigma(0)=-ig^2C(R) \int \frac{d^4
p}{(2\pi)^4} \frac{g_{\mu \nu}}{p^2} \gamma^{\mu}
\frac{p{\hspace{-1.9mm}\slash}+ K{\hspace{-3.1mm}\slash}}
{(p+K)^2} \gamma^{\nu} \eea and \bea \Sigma^{\prime}(K)=2g^2 C(R)
\int \frac{d^4 p}{(2\pi)^4} (p{\hspace{-1.9mm}\slash}+
K{\hspace{-3.1mm}\slash}) \left[
\frac{\Gamma_b(p)}{(p+K)^2}-\frac{\Gamma_f(p+K)}{p^2}+i
\Gamma_b(p) \Gamma_f(p) \right]. \eea If we take only the real
part $(\mbox{Re}\,\Sigma^{\prime}(K))$ of the contribution at
finite temperature and density we obtain \bea
\mbox{Re}\,\Sigma^{\prime}(K)=2g^2C(R) \int \frac{d^4 p}{(2\pi)^4}
\left[(p{\hspace{-1.9mm}\slash}+ K{\hspace{-3.1mm}\slash})
\Gamma_b(p)+ p{\hspace{-1.9mm}\slash} \Gamma_f(p) \right]
\frac{1}{(p+K)^2}. \la{rse} \eea Now we multiply $(\ref{rse})$ by
either $K{\hspace{-3.1mm}\slash}$ or $u{\hspace{-2.1mm}\slash}$,
then we take the trace and perform the integrations over $p_0$ and
the two angular variables, and finally we find that functions
$(\ref{lifa})$ and $(\ref{lifb})$ can be written as \bea
a(\omega,k)=g^2C(R)A(w,k,\mu_f), \la{aF0} \\
b(\omega,k)=g^2C(R)B(w,k,\mu_f), \la{bF0} \eea where we have used
the notation given in \cite{quimbay}. In the last expression,
$A(\omega,k,\mu_f)$ and $B(\omega,k,\mu_f)$ are integrals over the
modulus of the three-momentum $p= \vert \vec{p} \vert$ and they
are defined as \bea A(\omega,k,\mu_f) =
\frac{1}{k^2}\int^\infty_0\frac{dp}{8\pi^2} \left[ 2p-\frac{\omega
p}{k} \log \left( \frac{\omega+k}{\omega-k} \right) \right]
\left[2n_b(p)+n_f^-(p)+n_f^+(p) \right], \la{Alead} \eea \bea
B(\omega,k,\mu_f) = \frac{1}{k^2} \int^\infty_0\frac{dp}{8\pi^2}
\left[ \frac{p(\omega^2-k^2)}{k} \log \left(
\frac{\omega+k}{\omega-k} \right) -2\omega p \right]
\left[2n_b(p)+n_f^-(p)+n_f^+(p)\right]. \la{Blead} \nn \\
\eea The integrals $(\ref{Alead})$ and $(\ref{Blead})$ have been
obtained using the high density approximation, ${\it i. e.}$
$\mu_f \gg k$ and $\mu_f \gg \omega$, and keeping the leading
terms in temperature and chemical potential \cite{morales}.
Evaluating these integrals we obtain that $a(\omega,k)$ and
$b(\omega,k)$ are given by \bea a(\omega,k) &=& \frac{M_F^2}{k^2}
\left[ 1-\frac{\omega}{2k} \log
\frac{\omega+k}{\omega-k}\right], \la{a1} \\
b(\omega,k) &=& \frac{M_F^2}{k^2} \left[ \frac{\omega^2-k^2}{2k}
\log \frac{\omega+k}{\omega-k}-\omega \right], \la{b1} \eea where
fermion effective mass $M_F$ is \be M_F^2(T, \mu_f)=\frac{g^2
C(R)}{8} \left( T^2+\frac{\mu_f^2}{\pi^2} \right). \la{me} \ee The
value of $M_F$ given by $(\ref{me})$ is in agreement with
\cite{kajantie}-\cite{lebellac}. We are interested in the
effective mass at $T=0$, which corresponds precisely to the case
in which the vacuum is described by a medium at zero temperature.
For this case \be M_F^2 (0, \mu_F)= M_F^2=\frac{g^2 C(R)}{8}
\frac{\mu_f^2}{\pi^2}. \la{mef} \ee Substituting $(\ref{a1})$ and
$(\ref{b1})$ into $(\ref{fdr0})$, we obtain for the limit $k \ll
M_F$ that \be \omega^2(k) = M_F^2 \left[ 1 + \frac{2}{3}
\frac{k}{M_F} + \frac{5}{9} \frac{k^2}{M_F^2} + \dots \right]
\la{dr1}. \ee This dispersion relation is gauge invariant due to
that the calculation has been done at leading order in temperature
and chemical potential \cite{morales}.

It is well known that the relativistic energy in the vacuum for a
massive fermion at rest is $\omega^2 (0)= m_f^2$. From
$(\ref{dr1})$ we have that for $k=0$ then $\omega^2 (0) = M_F^2$
and thereby we can identify the fermion effective mass at zero
temperature as the rest mass of the fermion, i. e. $m_f^2=M_F^2$.
So the gauge invariant fermion mass which is generated by the
SU(N) gauge interaction of the massless fermion with the vacuum is
\be m_f^2 = \frac{g^2 C(R)}{8} \frac{\mu_f^2}{\pi^2}, \la{mas} \ee
where $\mu_f$ is a free parameter.


\subsection{Polarization tensor and gauge boson mass}
The gauge boson mass is due to the charge fluctuations of vacuum.
This mass is calculated following the general procedure presented
in the introduction. The most general form of the polarization
tensor which preserves invariance under rotations, translations
and gauge transformations is \cite{weldon2}

\be \Pi_{\mu \nu}(K) = P_{\mu \nu} \Pi_T (K) + Q_{\mu \nu} \Pi_L
(K), \la{pot} \ee where Lorentz-invariant functions $\Pi_L$ and
$\Pi_T$, which characterize the longitudinal and transverse modes
respectively, are obtained by contraction \bea
\Pi_L(K)=-\frac{K^2}{k^2}u^{\mu}u^{\nu}\Pi_{\mu \nu}, \la{potl} \\
\Pi_T(K)=-\frac{1}{2}\Pi_L + \frac{1}{2}g^{\mu \nu} \Pi_{\mu \nu}.
\la{pott} \eea Bosonic dispersion relations are obtained by
looking at the poles of the full propagator which results from
adding all vacuum polarization insertions. The full bosonic
propagator is \cite{weldon2} \be D_{\mu \nu}(K) = \frac{Q_{\mu
\nu}}{K^2 - \Pi_L (K)} + \frac{P_{\mu \nu}}{K^2 - \Pi_T (K)} -
(\xi - 1) \frac{K_\mu K_\nu}{K^4} , \la{fbp} \ee where $\xi$ is a
gauge parameter. The gauge invariant dispersion relations
describing the two propagation modes are found from \bea
K^2 - \Pi_L(K) = 0, \la{bdrl} \\
K^2 - \Pi_T(K) = 0. \la{bdrt} \eea The one-loop contribution to
vacuum polarization from the diagram which can be constructed in
this theory is given by \be \Pi_{\mu \nu} (K)= i g^2 C(R) \int
\frac{d^4p}{(2\pi )^4} Tr \left[ {\gamma }_\mu S(p) {\gamma }_\nu
S(p+K) \right],  \la{ptdf} \ee where $S$ is the fermion propagator
$(\ref{fp})$. Substituting $(\ref{fp})$ into $(\ref{ptdf})$ the
polarization tensor can be written as $\Pi_{\mu \nu}(K)= \Pi_{\mu
\nu}(0)+ \Pi'_{\mu \nu}(K)$, where $\Pi_{\mu \nu}(0)$ is the
contribution at zero density and temperature and $\Pi'_{\mu
\nu}(K)$ is the contribution at finite temperature and density.

The real part of the contribution at finite temperature and
density to the polarization tensor $\mbox{Re}\,\Pi'_{\mu \nu}(K)$
is given by \bea \mbox{Re}\,\Pi'_{\mu \nu}(K) = \frac{g^2 C(R)}{2}
\int \frac{d^4p}{\pi^4}\frac{(p^2 + p \cdot K) g^{\mu \nu} -
2p^{\mu} p^{\nu} - p^{\mu}K^{\nu} - p^{\nu}K^{\mu}}{(p+K)^2}
\Gamma_f(p). \la{rppt} \eea Substituting $(\ref{rppt})$ in
$(\ref{potl})$ and $(\ref{pott})$ and keeping the leading terms in
temperature and chemical potential we obtain that for the high
density approximation ($\mu_f \gg k$ and $\mu_f \gg \omega$)

\bea \mbox{Re}\,\Pi'_L(K)&&=3 M_B^2 \left[ 1 - \frac{\omega}{2k}
\log
\frac{\omega+k}{\omega-k} \right], \la{rplo} \\
\mbox{Re}\,\Pi'_T(K)&&=\frac{3}{2} M_B^2 \left[
\frac{\omega^2}{k^2} + \left( 1 - \frac{\omega^2}{k^2} \right)
\frac{\omega}{2k} \log \frac{\omega+k}{\omega-k} \right],
\la{rptr} \eea where the gauge boson effective mass $M_B$ is

\be M_B^2(T, \mu_f)= \frac{1}{6} N g^2 T^2 + \frac{1}{2} g^2 C(R)
\left[ \frac{T^2}{6} + \frac{\mu_f^2}{2 \pi^2} \right],
\la{effnab} \ee being $N$ the gauge group dimension. The
non-abelian effective mass $(\ref{effnab})$ is in agreement with
\cite{lebellac}. The abelian gauge boson associated with a U(1)
gauge invariant theory acquires an effective mass $M_{B(a)}$
defined by

\be M_{B(a)}^2(T, \mu_f)= e^2 \left[ \frac{T^2}{6} +
\frac{\mu_f^2}{2 \pi^2} \right], \la{effab} \ee where $e$ is the
interaction coupling constant associated with the U(1) abelian
gauge group. The abelian effective mass $(\ref{effab})$ is in
agreement with \cite{braaten}. Because the vacuum is described by
a virtual medium at $T=0$, then the non-abelian gauge boson
effective mass generated by quantum fluctuations of vacuum is

\be M_{B(na)}^2(0, \mu_f)= M_{B(na)}^2 = g^2 C(R)
\frac{\mu_f^2}{4\pi^2}, \la{bemT0} \ee and the abelian gauge boson
effective mass generated by quantum fluctuations of vacuum is

\be M_{B(a)}^2(0, \mu_f)= M_{B(a)}^2 = e^2 \frac{\mu_f^2}{2\pi^2},
\la{bemT0} \ee in agreement with the result obtained at finite
density and zero temperature \cite{altherr}. For the limit $k \ll
M_{B_\mu}$, we can obtain the dispersion relations for the
transverse and longitudinal propagation modes \cite{weldon2} \bea
\omega_L^2=M_{B}^2 + \frac{3}{5}k_L^2 + \ldots\la{disl} \\
\omega_T^2=M_{B}^2 + \frac{6}{5}k_T^2 + \ldots \la{dist} \eea We
note that $(\ref{disl})$ and $(\ref{dist})$ have the same value
when the three-momentum goes to zero. We van observe from
$(\ref{disl})$ and $(\ref{dist})$ that for $k=0$ then
$\omega^2(0)=M_{B}^2$ and we recognize the gauge boson effective
mass as a real gauge boson mass. The non-abelian gauge boson mass
is \be m_{b(na)}^2 = M_{B(na)}^2 = g^2 C(R)
\frac{\mu_f^2}{4\pi^2}, \la{nabosmas} \ee and the abelian gauge
boson mass is\be m_{b(a)}^2 = M_{B(a)}^2 = e^2
\frac{\mu_f^2}{2\pi^2}. \la{abosmas} \ee We observe that the gauge
boson mass is a function on the chemical potential that is a free
parameter on this approach. We note that if the fermionic chemical
potential has an imaginary value, then the gauge boson effective
mass is given by $(\ref{nabosmas})$ or $(\ref{abosmas})$, and it
would be negative \cite{bluhm}.


\section{SMWHS case}\label{sec:03}

In this section we follow the same procedure as the previous one.
We calculate fermion and electroweak gauge boson masses for the
case in which the dynamics of physical vacuum is described by the
SMWHS. The physical vacuum is assumed to be a medium at zero
temperature constituted by quarks, antiquarks, leptons and
antileptons interacting among themselves through gluons $G$ (for
the case of quarks and antiquarks), electroweak gauge bosons
$W^{\pm}$, non-abelian gauge bosons $W^3$ and abelian gauge bosons
$B$. In this quantum medium there exist an excess of virtual
antifermions over virtual fermions. This fact is described by
non-vanishing chemical potentials associated with different
fermion flavors. The chemical potentials for the six quarks are
represented by $\mu_u, \mu_d, \mu_c, \mu_s, \mu_t, \mu_b$. For the
chemical potentials of charged leptons we use $\mu_e, \mu_{\mu},
\mu_{\tau}$ and for neutrinos $\mu_{\nu_e},
\mu_{\nu_{\mu}},\mu_{\nu_{\tau}}$. These non-vanishing chemical
potentials are input parameters in the approach of particle mass
generation.

The dynamics of the vacuum associated with the strong interaction
is described by Quantum Chromodynamics (QCD), while the
electroweak dynamics of physical vacuum is described by the
$SU(2)_L \times U(1)_Y$ electroweak standard model without the
Higgs sector. This last model is defined by the following
Lagrangian density \be {\cal L}_{ew} = {\cal L}_{YM} + {\cal
L}_{FB} + {\cal L}_{GF} + {\cal L}_{FP}, \la{ldm} \ee where ${\cal
L}_{YM}$ is the Yang-Mills Lagrangian density, ${\cal L}_{FB}$ is
the fermionic-bosonic Lagrangian density, ${\cal L}_{GF}$ is the
gauge fixing Lagrangian density and ${\cal L}_{FP}$ is the
Fadeev-Popov Lagrangian density. The ${\cal L}_{YM}$ is given by
\be {\cal L}_{YM} = -\frac{1}{4} W_{A}^{\mu \nu} W_{\mu \nu}^{A}
-\frac{1}{4} F^{\mu \nu} F_{\mu \nu}, \la{lym} \ee where $W_{\mu
\nu}^{A} =\partial_{\mu} W_{\nu}^A - \partial_{\mu} W_{\mu}^A +
g_w F^{ABC} W_{\mu}^B W_{\nu}^C $ is the energy-momentum tensor
associated with the group $SU(2)_L$ and $ F_{\mu \nu} =
\partial_{\mu}B_{\nu} - \partial_{\mu} B_{\mu}$ is the one
associated with the group $U(1)_Y$ . The ${\cal L}_{FB}$ is
written as \be {\cal L}_{FB} = i \bar{\mbox{L}}
\gamma^{\mu}\mbox{D}_\mu \mbox{L} + i \psi^i_R
\gamma^{\mu}\mbox{D}_\mu \psi^i_R + i \psi^I_R
\gamma^{\mu}\mbox{D}_\mu \psi^I_R, \la{lafb} \ee where
$\mbox{D}_\mu \mbox{L} = ( \partial_{\mu} + ig_e Y_L B_\mu/2 +
ig_w T_i W_{\mu}^i ) \mbox{L}$ and $\mbox{D}_\mu \mbox{R} =
(\partial_{\mu} + ig_e Y_R B_\mu/2 ) \mbox{R}$, being $g_w$ the
gauge coupling constant associated with the group $SU(2)_L$ ,
$g_e$ the one associated with the group $U(1)_Y$, $Y_L = -1$, $Y_R
= -2$ and $T_i = \sigma_i /2$. The $SU(2)_L$ left-handed doublet
$(\mbox{L})$ is given by \be \mbox{L}= {\psi^i \choose {\psi^I}}_L
. \ee


\subsection{Masses of fermions}

Initially we consider the propagation of massless fermions in a
medium at finite temperature and density. The fermion masses are
calculated following the same procedure as on previous section.
For a non-abelian gauge theory with parity violation and chirality
conservation like the SMWHS, the real part of the self-energy for
a massless fermion is \cite{quimbay} \be \mbox{Re}\,\Sigma'(K)=-
K{\hspace{-3.1mm}\slash}(a_{L}P_L +a_{R}P_R)-
u{\hspace{-2.2mm}\slash}(b_{L}P_L +b_{R}P_R), \ee where $P_L
\equiv\frac{1}{2}(1-\gamma_5)$ and $P_R
\equiv\frac{1}{2}(1+\gamma_5)$ are the left- and right-handed
chiral projectors respectively. The functions $a_L$, $a_R$, $b_L$
and $b_R$ are the chiral projections of Lorentz-invariant
functions $a$, $b$ and they are defined as follows \bea
a &=& a_L P_L + a_R P_R, \\
b &=& b_L P_L + b_R P_R. \eea The inverse fermion propagator is
given by \be S^{-1}(K)= {\cal L}{\hspace{-2.5mm}\slash} P_L +
\Re{\hspace{-2.5mm}\slash} P_R, \la{ifp} \ee where \bea
{\cal L}^{\mu} &=& ( 1 + a_L) K^{\mu} + b_L u^{\mu}, \\
{\Re}^{\mu} &=& ( 1 + a_R) K^{\mu} + b_R u^{\mu}. \eea The fermion
propagator follows from the inversion of $(\ref{ifp})$ \bea
S=\frac{1}{D}\left[\left({\cal L}^2\Re{\hspace{-2.5mm}
\slash}\right)P_L + \left(\Re^2{\cal L}{\hspace{-2.5mm}\slash}
\right)P_R \right],\la{p1} \eea where $D(\omega,k)={\cal L}^2
{\Re}^2$. The poles of the propagator correspond to values
$\omega$ and $k$ for which the determinant $D$ in (\ref{p1})
vanishes \be {\cal L}^2 {\Re}^2 =0.\la{d} \ee In the rest frame of
the dense plasma $u=(1,\vec 0)$, Eq.$(\ref{d})$ leads to fermion
dispersion relations for a chirally invariant gauge theory with
parity violation, as from the case of the SMWHS. Thus fermion
dispersion relations are given by \cite{quimbay} \bea \left[
\omega (1+a_L)+b_L \right]^2- k^2 \left[ 1+a_L \right]^2 &=& 0,
\la{dra}\\
\left[ \omega (1+a_R)+b_R \right]^2-k^2 \left[ 1+a_R \right]^2 &=&
0. \la{drb} \eea Left- and right-handed components of the fermion
dispersion relations are decoupled relations. The Lorentz
invariant functions $a(\omega,k)$ and $b(\omega,k)$ are calculated
from expressions $(\ref{lifa})$ and $(\ref{lifb})$ through the
real part of fermion self-energy. This self-energy is obtained
adding all posible gauge boson contributions admitted by the
Feynman rules of the SMWHS.

We work on the basis of gauge bosons given by $B_\mu$,
$W_{\mu}^3$, $W_{\mu}^{\pm}$, where the charged electroweak gauge
bosons are $W_{\mu}^{\pm} = (W_{\mu}^1 \mp iW_{\mu}^2)/ \sqrt{2}$.
The diagrams with an exchange of a $W^{\pm}$ gauge boson induce a
flavor change in the incoming fermion $i$ to a different outgoing
fermion $j$.


\subsubsection{Quark masses}

The quark masses are obtained from flavor change contributions
previously mentioned. For the quark sector, the flavor $i$ $(I)$
of the internal quark (inside the loop) runs over the up $(i)$ or
down $(I)$ quark flavors according to the type of the external
quark (outside the loop). As for each contribution to the quark
self-energy is proportional to $(\ref{Alead})$-$(\ref{Blead})$,
the functions $a_L$, $a_R$, $b_L$ and $b_R$ are given by \bea
a_L(\omega,k)_{ij} &=& [f_{S}+f_{W^3}+f_{B}] A(\omega,k,\mu_{i}) +
\sum_{I}
f_{W^{\pm}} A(\omega,k,\mu_{I}), \la{al} \\
b_L(\omega,k)_{ij} &=& [f_{S}+f_{W^3}+f_{B}] B(\omega,k,\mu_{i}) +
\sum_{I}
f_{W^{\pm}} B(\omega,k,\mu_{I}), \la{bl} \\
a_R(\omega,k)_{ij} &=&
[f_{S}+f_{B}] A(\omega,k,\mu_{i}), \la{ar} \\
b_R(\omega,k)_{ij} &=& [f_{S}+f_{B}] B(\omega,k,\mu_{i}). \la{br}
\eea In the last expressions the coefficients $f$ are \bea
f_{S} &=& \frac{4}{3}g_s^2 \delta_{ij}, \la{cfs}\\
f_{W^3}&=& \frac{1}{4} g_w^2 \delta_{ij}, \la{cfw3} \\
f_{B} &=& \frac{1}{4} g_e^2 \delta_{ij}, \la{cfb} \\
f_{W^{\pm}} &=& \frac{1}{2} g_w^2 K_{il}^{+}K_{lj},\la{cfwmm} \eea
where ${\it K}$ represents the CKM matrix and $g_s$ is the strong
running coupling constant associated with the group $SU(3)_C$. The
integrals $A(\omega,k,\mu_f)$ and $B(\omega,k,\mu_f)$ are obtained
in a high density approximation $(\mu_f \gg k$ and $\mu_f \gg
\omega )$. These integrals keeping the leading terms in
temperature and chemical potential are given by \bea
A(\omega,k,\mu_f) &=& \frac{1}{8 k^2} \left( T^2+ \frac{\mu_f^2}
{\pi^2} \right)\left[1-\frac{\omega}{2k} \log \frac{\omega+k}
{\omega-k} \right] \la{ALR},
\\
B(\omega,k,\mu_f) &=& \frac{1}{8 k^2} \left( T^2+ \frac{\mu_f^2}
{\pi^2}\right)\left[\frac{\omega^2-k^2}{2k} \log \frac{\omega+k}
{\omega-k}-\omega \right].\la{BLR} \eea

The chiral projections of the Lorentz-invariant functions are \bea
a_{L}(\omega,k)_{ij} &=& \frac{1}{8k^2}\left[1-F(\frac{\omega}{k})
\right]\left[l_{ij}(T^2+\frac{\mu_i^2}{\pi^2})+ h_{ij}(T^2+
\frac{\mu_i^2}{\pi^2})\right], \la{aL} \\
b_{L}(\omega,k)_{ij} &=& -\frac{1}{8k^2}\left[\frac{\omega}{k}+
(\frac{k}{\omega}-\frac{\omega}{k})F(\frac{\omega}{k})\right]
\left [l_{ij}(T^2+\frac{\mu_i^2}{\pi^2})+
h_{ij}(T^2+\frac{\mu_i^2}{\pi^2})
\right], \nn \la{bL} \\
\\
a_{R}(\omega,k)_{ij} &=& \frac{1}{8k^2}\left[1-F(\frac{\omega}{k})
\right]\left[r_{ij}(T^2+\frac{\mu_i^2}{\pi^2}) \right], \la{aR} \\
b_{R}(\omega,k)_{ij} &=& -\frac{1}{8k^2}\left[\frac{\omega}{k}+
(\frac{k}{\omega}-\frac{\omega}{k})F(\frac{\omega}{k})\right]
\left[r_{ij}(T^2+\frac{\mu_i^2}{\pi^2})\right], \la{bR} \eea where
$F(x)$ is \be F(x)=\frac{x}{2} \log \left(\frac{x+1}{x-1}, \right)
\ee and the coefficients $l_{ij}$, $h_{ij}$ and $r_{ij}$ are given
by \bea l_{ij} &=& \left( \frac{4}{3}g_s^2 + \frac{1}{4}g_w^2 +
\frac{1}{4}g_e^2 \right)\delta_{ij}, \\
h_{ij} &=& \sum_l \left(\frac{g_w^2}{2}\right) K_{il}^+ K_{lj}, \\
r_{ij} &=& \left(\frac{4}{3}g_s^2 + \frac{1}{4}g_e^2 \right)
\delta_{ij}. \eea Substituting $(\ref{aL})$-$(\ref{bL})$ into
$(\ref{dra})$, and $(\ref{aR})$-$(\ref{bR})$ into $(\ref{drb})$,
for the limit $k \ll M_{(i,I)_{L,R}}$ we obtain \be \omega^2(k) =
M_{(i,I)_{L,R}}^2 \left[ 1 + \frac{2}{3} \frac{k}{M_{(i,I)_{L,R}}}
+ \frac{5}{9} \frac{k^2}{M_{(i,I)_{L,R}}^2} + \dots \right],
\la{drsmlr} \ee where \bea M_{(i,I)_L}^2(T,\mu_f) &=&
(l_{ij}+h_{ij}) \frac{T^2}{8} + l_{ij} \frac{\mu_{(i,I)_L}^2}{8
\pi^2} + h_{ij} \frac{\mu_{(I,i)_L}^2}
{8 \pi^2}, \la{emLq}\\
M_{(i,I)_R}^2(T,\mu_f)  &=& r_{ij} \frac{T^2}{8} + r_{ij}
\frac{\mu_{(i,I)_R}^2}{8 \pi^2}. \la{emRq} \eea As it was
explained on previous section, we are interested in effective
masses at $T=0$. For this case \bea M_{(i,I)_L}^2(0,\mu_f) &=&
l_{ij} \frac{\mu_{(i,I)_L}^2}{8 \pi^2} +
h_{ij} \frac{\mu_{(I,i)_L}^2}{8 \pi^2}, \la{emLT0}\\
M_{(i,I)_R}^2(0,\mu_f)  &=& r_{ij} \frac{\mu_{(i,I)_R}^2}{8
\pi^2}. \la{emRT0} \eea

Keeping the same argument as in section 2, we can identify quark
effective masses at zero temperature with the rest masses of
quarks. Coming from the left-handed and right-handed
representations, we find that masses of the left-handed quarks are
\bea m_{i}^2 = \left[ \frac{4}{3}g_s^2 + \frac{1}{4}g_w^2 +
\frac{1}{4} g_e^2 \right] \frac{\mu_{i_L}^2}{8 \pi^2} + \left[
\frac{1}{2}g_w^2
\right] \frac{\mu_{I_L}^2}{8 \pi^2}, \la{uqpl} \\
m_{I}^2 = \left[ \frac{4}{3}g_s^2 + \frac{1}{4}g_w^2 + \frac{1}{4}
g_e^2 \right] \frac{\mu_{I_L}^2}{8 \pi^2} + \left[
\frac{1}{2}g_w^2 \right] \frac{\mu_{i_L}^2}{8 \pi^2},  \la{dqpl}
\eea and the masses of the right-handed quarks are \bea m_{i}^2 =
\left[ \frac{4}{3}g_s^2 + \frac{1}{4} g_e^2 \right]
\frac{\mu_{i_R}^2}{8 \pi^2}, \la{uqpr} \\
m_{I}^2 = \left[ \frac{4}{3}g_s^2 + \frac{1}{4} g_e^2 \right]
\frac{\mu_{I_R}^2}{8 \pi^2}, \la{dqpr} \eea where the couple of
indexes $(i,I)$ run over quarks $(u, d)$, $(c, s)$ and $(t, b)$.
It is known that the masses of left-handed quarks are the same as
the masses of right-handed quarks. This means that left-handed
quark chemical potentials $\mu_{q_L}$ are different from
right-handed quark chemical potentials $\mu_{f_R}$.

If we call \bea a_q &=&\frac{1}{8 \pi^2} \left[ \frac{4}{3}g_s^2 +
\frac{1}{4}g_w^2 +\frac{1}{4} g_e^2 \right], \la{caq}\\
b_q &=&\frac{1}{8 \pi^2} \left[ \frac{1}{2}g_w^2 \right], \la{cbq}
\eea the expressions $(\ref{uqpl})$ and $(\ref{dqpl})$ lead us to
\bea
\mu_{u_L}^2 &=& \frac{a_q m_u^2 - b_q m_d^2}{a_q^2 - b_q^2}, \la{pqu2} \\
\mu_{d_L}^2 &=& \frac{-b_q m_u^2 + a_q m_d^2}{a_q^2 - b_q^2}.
\la{pqd2} \eea Naming \be c_q =\frac{1}{8 \pi^2} \left[
\frac{4}{3}g_s^2 + \frac{1}{4} g_e^2 \right], \la{ccq} \ee the
expressions $(\ref{uqpr})$ and $(\ref{dqpr})$ can be written as
\bea
\mu_{u_R}^2 &=& \frac{m_u^2}{c_q}, \la{pquR2} \\
\mu_{d_R}^2 &=& \frac{m_u^2}{c_q}, \la{pqdR2} \eea and similar
expressions for the other two quark doublets $(c, s)$ and $(t,
b)$.

If we take the experimental central values for the strong constant
as $\alpha_s=0.1184$, the fine-structure constant as
$\alpha_e=7.2973525376 \times 10^{-3}$ and the cosine of the
electroweak mixing angle as $\cos \theta_w =
M_W/M_Z=80.399/91.1876=0.88168786$ \cite{pdg}, then $g_s=1.21978$,
$g_w=0.641799$ and $g_e=0.343457$. Fixing the central values for
quark masses as \cite{pdg} $m_u = 0.0025$ GeV, $m_d = 0.00495$
GeV, $m_c = 1.27$ GeV, $m_s = 0.101$ GeV, $m_t = 172.0$ GeV, $m_b
= 4.19$ GeV,  into the expressions $(\ref{pqu2})$, $(\ref{pqd2})$,
$(\ref{pquR2})$ and $(\ref{pqdR2})$, we obtain that the squares of
left-handed quark chemical potentials are given by \bea
\mu_{u_L}^2 &=& 1.4559 \times
10^{-4}, \la{vpqu2} \\ \mu_{d_L}^2 &=& 9.0 \times 10^{-4},\\
\mu_{c_L}^2 &=& 60.7141, \\
\mu_{s_L}^2 &=& -5.5280, \\ \mu_{t_L}^2 &=& 1.1143 \times 10^{6}, \\
\mu_{b_L}^2 &=& -1.0778 \times 10^{5}, \la{vpqb2}\eea and the
squares of right-handed quark chemical potentials are \bea
\mu_{u_R}^2 &=& 2.4511 \times
10^{-4}, \la{vpqu2} \\ \mu_{d_R}^2 &=& 9.6092 \times 10^{-4},\\
\mu_{c_R}^2 &=& 63.254, \\
\mu_{s_R}^2 &=& 0.4, \\ \mu_{t_R}^2 &=& 1.1602 \times 10^{6}, \\
\mu_{b_R}^2 &=& 688.508, \la{vpqb2}\eea where the left- and
right-handed chemical potentials are given in GeV$^2$ units.


\subsubsection{Lepton masses}

For the lepton sector, the contributions to the fermion
self-energy are proportional to $(\ref{Alead})$-$(\ref{Blead})$
and functions $a_L$, $a_R$, $b_L$ and $b_R$ are given by \bea
a_L(\omega,k)_{ij} &=& [f_{W^3}+f_{B}] A(\omega,k,\mu_{i}) +
\sum_{I}
f_{W^{\pm}} A(\omega,k,\mu_{I}), \la{al} \\
b_L(\omega,k)_{ij} &=& [f_{W^3}+f_{B}] B(\omega,k,\mu_{i}) +
\sum_{I}
f_{W^{\pm}} B(\omega,k,\mu_{I}), \la{bl} \\
a_R(\omega,k)_{ij} &=& [f_{B}] A(\omega,k,\mu_{i}), \la{ar} \\
b_R(\omega,k)_{ij} &=& [f_{B}] B(\omega,k,\mu_{i}), \la{br} \eea
where $f_{W^{\pm}}= g_w^2/2$ and the other coefficients $f_{W^3}$
and $f_{B}$ are given by $(\ref{cfw3})$ and $(\ref{cfb})$
respectively.

The dispersion relation for leptons are similar to the relations
$(\ref{drsmlr})$, but in this case the effective masses
$(\ref{emLq})$  and $(\ref{emRq})$ are given by \bea
M_{(i,I)_L}^2(T,\mu_f) &=& (l+h) \frac{T^2}{8} + l
\frac{\mu_{(i,I)_L}^2}{8 \pi^2} + h \frac{\mu_{(I,i)_L}^2}
{8 \pi^2}, \la{emLl}\\
M_{(i,I)_R}^2(T,\mu_f)  &=& r \frac{T^2}{8} + r
\frac{\mu_{(i,I)_R}^2}{8 \pi^2}, \la{emRl} \eea where the
coefficients $l$, $h$ and $r$ for the charged leptons are given by
\bea
l &=& \left( \frac{1}{4}g_w^2 + \frac{1}{4}g_e^2 \right), \\
h &=& \left( \frac{1}{2} g_w^2 \right), \\
r &=& \left(\frac{1}{4}g_e^2 \right), \eea and for the neutrinos
these coefficients are \bea
l &=& \left( \frac{1}{4}g_w^2 \right), \\
h &=& \left( \frac{1}{2} g_w^2 \right), \\
r &=& 0. \eea

The leptonic effective masses $(\ref{emLl})$  and $(\ref{emRl})$
at zero temperature can be interpreted as lepton masses. Coming
from left-handed and right-handed representations, we find that
masses of left-handed leptons are given by \bea m_{i}^2 = \left[
\frac{1}{4}g_w^2 \right] \frac{\mu_{i_L}^2}{8 \pi^2} + \left[
\frac{1}{2}g_w^2
\right] \frac{\mu_{I_L}^2}{8 \pi^2}, \la{nlpl} \\
m_{I}^2 = \left[ \frac{1}{4}g_w^2 + \frac{1}{4} g_e^2 \right]
\frac{\mu_{I_L}^2}{8 \pi^2} + \left[ \frac{1}{2}g_w^2 \right]
\frac{\mu_{i_L}^2}{8 \pi^2},  \la{elpl} \eea and masses of the
right-handed charged leptons are \bea m_{I}^2 = \left[ \frac{1}{4}
g_e^2 \right] \frac{\mu_{I_R}^2}{8 \pi^2}, \la{elpr} \eea where
couple of indexes $(i,I)$ run over leptons $(\nu_e, e)$,
$(\nu_{\mu}, \mu)$ and $(\nu_{\tau}, \tau)$. We observe that the
approach predicts that neutrinos are massive. The $W^3$ and
$W^{\pm}$ interactions among massless neutrinos with physical
vacuum are the source for left-handed neutrinos masses, as we can
observe from $(\ref{nlpl})$.

If we call \bea a_l&=&\frac{1}{8 \pi^2} \left[ \frac{1}{4}g_w^2
\right], \la{cal}\\
b_l &=&\frac{1}{8 \pi^2} \left[ \frac{1}{2}g_w^2 \right], \la{cbl} \\
c_l &=&\frac{1}{8 \pi^2} \left[ \frac{1}{4}g_w^2+\frac{1}{4}g_e^2
\right], \la{cbl} \eea then the expressions $(\ref{nlpl})$ and
$(\ref{elpl})$ lead us to \bea \mu_{\nu_L}^2 &=& \frac{c_l m_\nu^2
- b_l m_e^2}{a_l c_l - b_l^2},
\la{pqn2} \\
\mu_{e_L}^2 &=& \frac{-b_l m_\nu^2 + a_l m_e^2}{a_l c_l - b_l^2}.
\la{pqe2} \eea Calling \be d_l =\frac{1}{8 \pi^2} \left[
\frac{1}{4} g_e^2 \right], \la{ccq} \ee the expression
$(\ref{elpr})$ can be written as \be \mu_{e_R}^2 =
\frac{m_e^2}{d_l}, \la{pqeR2} \ee and similar expressions for the
other two lepton doublets $(\nu_\mu, \mu)$ and $(\nu_\tau, \tau)$.
Assuming that neutrinos are massless $m_{\nu_e} = m_{\nu_\mu}
=m_{\nu_\tau} =0$, and fixing the experimental central values for
charged lepton masses as \cite{pdg} $m_e = 0.510998910 \times
10^{-3}$ GeV, $m_\mu= 0.105658367$ GeV, $m_\tau = 1.77682$ GeV,
into the expressions $(\ref{pqn2})$, $(\ref{pqe2})$ and
$(\ref{pqeR2})$, we obtain that the squares of left-handed lepton
chemical potentials are \bea \mu_{\nu_{e_L}}^2 &=& 1.4699 \times
10^{-4}, \la{vpqn2}
\\ \mu_{e_L}^2 &=& -7.3492 \times
10^{-5}, \\ \mu_{\nu_{\mu_L}}^2 &=& 6.30867, \\ \mu_{\mu_L}^2 &=&
-3.1543, \\
\mu_{\nu_{\tau_L}}^2 &=& 1.7841 \times 10^{4}\\ \mu_{\tau_L}^2 &=&
-8.9205 \times 10^{3}, \la{vpqtau2} \eea and the squares of
right-handed charged lepton chemical potentials are \bea
\mu_{e_R}^2 &=& 6.9637 \times
10^{-4}, \la{vpqe2} \\ \mu_{\mu_R}^2 &=& 29.8888, \\
\mu_{\tau_R}^2 &=& 8.4526 \times 10^{3}, \la{vpqtau2}\eea where
the left- and right-handed chemical potentials are given in
GeV$^2$ units. Experimental neutrino masses are unknown but
experimental results show that neutrino masses are of order $1$ eV
\cite{pdg}, and cosmological interpretations from five-year WMAP
observations find a limit over the total mass of massive neutrinos
of $ \Sigma m_\nu < 0.6$ eV ($95\%$ CL) \cite{wmap5year}. These
results assure that values of left-handed lepton chemical
potentials obtained by taking neutrinos to be massless will change
a little if we have the real small neutrinos masses.

We observe that for five from the six fermion doublets the square
of the chemical potential associated to the down fermion of the
doublet has a negative value. This behavior is observed if there
is a large difference between the masses of the two fermions of
the doublet. This means that for the quark doublet which is formed
by the up and down quarks this behavior is not observed due to
that the masses for these two quarks are quite near. In this case,
the chemical potentials associated to these two quarks are
positive.

From expressions $(\ref{uqpl})$, $(\ref{dqpl})$, $(\ref{nlpl})$
and $(\ref{elpl})$ it can be seen that our approach does not
predict fermion mass values owing to the fermionic chemicals
potentials $\mu_{f_i}$ are free parameters. However, we have fixed
the values of these $\mu_{f_i}$ starting from the known
experimental values for fermion masses. This limitation of our
approach is similar to what happens in the MSM with Higgs
mechanism in the sense that fermion masses depend on the Yukawa
coupling constants which are free parameters. Similarly to occur
here when we find the values of vacuum chemical potentials, the
Yukawa coupling constants can be fixed by means of the
experimental values of fermion masses.


\subsection{Masses of electroweak gauge bosons}

The masses of electroweak gauge bosons are originated from charge
fluctuations of vacuum. These masses are calculated following a
sequence of steps that we present now: On the outset we write the
one-loop polarization tensor at finite temperature and density,
then we calculate the one-loop bosonic dispersion relations in the
high density approximation by obtaining the poles of gauge boson
propagators, next from these dispersion relations we obtain the
electroweak gauge boson effective masses at finite temperature and
density, finally we identify these effective masses at zero
temperature with masses of the electroweak gauge bosons.

To evaluate the bosonic polarization tensor associated with the
$W^{\pm}_\mu$, $W^3_\mu$, $B_\mu$ gauge boson propagators, we
follow the same procedure as in section 2.2. Applying the
expressions $(\ref{nabosmas})$ and $(\ref{abosmas})$ in the SMWHS
we obtain that the masses of the gauge bosons are

\bea M_{W^\pm}^2 &=& \frac{g_w^2}{2} \,\ \frac{S(\mu_{q_L}^2)+
\sum_{i=1}^{3} (\mu_{\nu_{i_L}}^2 -
\mu_{e_{i_L}}^2 )}{4\pi^2}, \la{mw+-} \\
M_{W^3}^2 &=& \frac{g_w^2}{4} \,\ \frac{S(\mu_{q_L}^2)+
\sum_{i=1}^{3} (\mu_{\nu_{i_L}}^2 -
\mu_{e_{i_L}}^2 )}{2\pi^2}, \la{mw3} \\
M_{B}^2 &=& \frac{g_e^2}{4} \,\ \frac{S(\mu_{q_L}^2)+
\sum_{i=1}^{3} (\mu_{\nu_{i_L}}^2 - \mu_{e_{i_L}}^2 )}{2\pi^2},
\la{mb} \eea where $S(\mu_{q_L}^2)=\mu_{u_L}^2 + \mu_{d_L}^2
+\mu_{c_L}^2 - \mu_{s_L}^2+\mu_{t_L}^2 - \mu_{b_L}^2$ and the sum
runs over the three lepton families. We remind you that if the
left-handed fermionic chemical potential has an imaginary value,
then its contribution to the gauge boson effective mass, as in the
case $(\ref{nabosmas})$ or $(\ref{abosmas})$, would be negative.
This fact means that finally the contribution from each fermionic
chemical potential to gauge boson masses is always positive.

Substituting the obtained left-handed fermionic chemical potential
values $(\ref{vpqu2})$-$(\ref{vpqb2})$ and
$(\ref{vpqn2})$-$(\ref{vpqtau2})$ into the expressions
$(\ref{mw+-})$-$(\ref{mb})$, we obtain \bea
M_{W^{\pm}}= M_{W^{3}}= 79.9344 \,{\mbox GeV}, \la{masw1} \\
M_{B}= 42.7767 \,{\mbox GeV}, \la{masb} \eea We observe that the
value of $M_W$ is smaller respect to its experimental value given
by $M_W^{exp}= 80.399 \pm 0.023$ GeV \cite{pdg}.

Due to well known physical reasons $W_{\mu}^3$ and $B_\mu$ gauge
bosons are mixed. After diagonalization of the mass matrix, we
obtain that the physical fields $A_\mu$ and $Z_\mu$ corresponding
to massless photon and neutral $Z^0$ boson of mass $M_Z$
respectively are related by means of \cite{weinberg1, pestieau}

\bea
M_Z^2 = M_W^2 + M_B^2 , \la{masz} \\
\cos \theta_w = \frac{M_W}{M_Z} \hspace{3.0mm} ,  \hspace{3.0mm}
\sin \theta_w = \frac{M_B}{M_Z}, \la{mix} \eea where $\theta_w$ is
the weak mixing angle \bea
Z_{\mu}^0 = B_\mu \sin \theta_w - W_{\mu}^3 \cos \theta_w , \la{zbw} \\
A_{\mu} = B_\mu \cos \theta_w + W_{\mu}^3 \sin \theta_w . \la{abw}
\eea Substituting $(\ref{masw1})$ and $(\ref{masb})$ into
$(\ref{masz})$ we obtain \be M_{Z}= 90.6606 \,{\mbox GeV},
\la{masZ0} \ee which is also smaller respect to its experimental
value given by $M_Z^{exp}= 91.1876 \pm 0.0021$ GeV \cite{pdg}.

Substituting the expressions for the fermionic chemical potentials
given by $(\ref{pqu2})$, $(\ref{pqd2})$, $(\ref{pqn2})$,
$(\ref{pqe2})$ into the expressions $(\ref{mw+-})$, $(\ref{mw3})$,
$(\ref{mb})$ we obtain that the masses of the electroweak gauge
bosons $W$ and $Z$ are given by

\bea
M_W^2 &=& g_w^2 (A_1 + A_2 +A_3-A_4) , \la{mw2mfrc} \\
M_Z^2 &=& (g_e ^2 + g_w^2) (A_1 + A_2 +A_3-A_4), \la{mz2mfrc} \eea
where the parameters $A_1$, $A_2$, $A_3$ and $A_4$ are \bea
A_1 &=& \frac{m_u^2+m_d^2}{B_1}, \la{A1} \\
A_2 &=& \frac{m_c^2-m_s^2+m_t^2-m_b^2}{B_2}, \la{A2} \\
A_3 &=&
\frac{3(m_e^2+m_\mu^2+m_\tau^2)}{B_3}, \la{A3} \\
A_4 &=& \frac{(3+g_e^2/g_w^2)(m_{\nu_e}^2 +m_{\nu_\mu}^2
+m_{\nu_\tau}^2)}{B_3}, \la{A4} \eea being \bea
B_1 &=& \frac{4}{3}g_s^2 + \frac{3}{4}g_w^2 + \frac{1}{4}g_e^2, \la{B1} \\
B_2 &=& \frac{4}{3}g_s^2 - \frac{1}{4}g_w^2 + \frac{1}{4}g_e^2, \la{B2} \\
B_3 &=& \frac{3}{4}g_w^2 - \frac{1}{4}g_e^2. \la{B3} \eea We
observe that $M_W$ and $M_Z$ are written in terms of the masses of
fermions and running coupling constants of the strong, weak and
electromagnetic interactions.

If we take the experimental central values for the strong running
coupling constant as $\alpha_s=0.1184$, the fine-structure
constant as $\alpha_e=7.2973525376 \times 10^{-3}$ and the cosine
of the electroweak mixing angle as $\cos \theta_w =
M_W/M_Z=80.399/91.1876=0.88168786$  \cite{pdg}, then
$g_s=1.21978$, $g_w=0.641799$ and $g_e=0.343457$. Substituting the
values of $g_s$, $g_w$ and $g_e$ and the values for the
experimental masses of the electrically charged fermions, given by
\cite{pdg} $m_u = 0.0025$ GeV, $m_d = 0.00495$ GeV, $m_c = 1.27$
GeV, $m_s = 0.101$ GeV, $m_t = 173.0015$ GeV, $m_b = 4.19$ GeV,
$m_e = 0.510998910 \times 10^{-3}$ GeV, $m_\mu= 0.105658367$ GeV,
$m_\tau = 1.77682$ GeV, into the expressions $(\ref{mw2mfrc})$ and
$(\ref{mz2mfrc})$, and assuming neutrinos as massless particles,
$m_{\nu_e} = m_{\nu_\mu} =m_{\nu_\tau} =0$, we obtain that
theoretical masses of the $W$ and $Z$ electroweak gauge bosons are
given by

\bea M_{W^{\pm}}^{th} &=& 79.9344 \pm 1.0208 \,{\mbox GeV},
\\ M_{Z}^{th} &=& 90.6606 \pm 1.1587 \,{\mbox GeV}. \la{masZ0} \la{masw}
\eea These theoretical masses are in agreement with theirs
experimental values given by $M_W^{exp}= 80.399 \pm 0.023$ GeV and
$M_Z^{exp}= 91.1876 \pm 0.0021$ GeV \cite{pdg}. Central values for
parameters $A_1$, $A_2$, $A_3$ and $A_4$ in expressions
$(\ref{mw2mfrc})$ and $(\ref{mz2mfrc})$ are $A_1=1.32427 \times
10^{-5}$, $A_2=15478$, $A_3=34.0137$ and $A_4=0$. We observe that
$A_2$ is very large respect to $A_3$ and $A_1$. Taking into
account the definition of parameter $A_2$ given by $(\ref{A2})$ we
can conclude that masses of electroweak gauge bosons coming
specially from top quark mass $m_t$ and strong running coupling
constant $g_s$.

We obtain also a prediction for top quark mass starting from the
expression $(\ref{mz2mfrc})$. Using central experimental values
$M_W^{exp}= 80.399$ GeV, $M_Z^{exp}= 91.1876$ GeV and considering
the uncertainties for running coupling constants and for fermion
masses, and assuming neutrinos as massless particles, we predict
from $(\ref{mz2mfrc})$ that top quark mass is $m_t^{th}=173.0015
\pm 0.6760$ GeV. This theoretical value is in agreement with the
experimental value for top quark mass given by \cite{pdg}
$m_t^{exp}=172.0 \pm 2.2$ GeV.


\section{Some relations between $m_t$ and $M_W$, $M_Z$}\label{sec:04}

The square of the electroweak gauge boson masses $M_W$ and $M_Z$
were written in terms of the fermion masses and the running
coupling constants of strong, weak and electromagnetic
interactions, such as shown in $(\ref{mw2mfrc})$ and
$(\ref{mz2mfrc})$. If we sum up $(\ref{mw2mfrc})$ and
$(\ref{mz2mfrc})$ we can write that

\be M_W^2 + M_Z^2 = (g_e ^2 + 2g_w^2) (A_1 + A_2 +A_3-A_4).
\la{mw2mz2} \ee Since the top quark mass $m_t$ is very large in
comparison to other fermion masses, it is very easy to prove that

\be A_1 + A_2 +A_3-A_4 \approx \frac{m_t^2}{B_2}. \la{A1A2A3A4}
\ee Substituting $(\ref{A1A2A3A4})$ into $(\ref{mw2mz2})$ we can
obtain that the square of the top quark mass and the squares of
the electroweak gauge boson masses are related as

\be m_t^2 = C_1 (M_W^2 + M_Z^2), \la{mt2mw2mz2} \ee where \be C_1
= \frac{B_2}{g_e^2+2g_w^2}. \la{defC1} \ee

On the other hand, if we take the square root of expressions
$(\ref{mw2mfrc})$ and $(\ref{mz2mfrc})$ we can prove that

\be (M_W + M_Z)^2 = (g_w + \sqrt{g_w^2 + g_e^2})^2 (A_1 + A_2
+A_3-A_4). \la{r2mw2mz2} \ee Substituting $(\ref{A1A2A3A4})$ into
$(\ref{r2mw2mz2})$ and after taking the square root, we can obtain
that the top quark mass and the electroweak gauge boson masses
satisfy the following relation

\be m_t = C_2 (M_W + M_Z) \approx M_W + M_Z, \la{mtmwmz} \ee where
\be C_2 = \frac{\sqrt{B_2}}{g_w + \sqrt{g_w^2 + g_e^2}}.
\la{defC2} \ee

Substituting the values $g_s=1.21978$, $g_w=0.641799$ and
$g_e=0.343457$ into $(\ref{defC1})$ and $(\ref{defC2})$ we obtain
that $C_1=2.02843$ and $C_2=1.00907$. Using the central values for
the electroweak gauge boson masses $M_W^{exp}= 80.399$ GeV and
$M_Z^{exp}= 91.1876$ GeV, from $(\ref{mt2mw2mz2})$ and
$(\ref{mtmwmz})$ we obtain that $m_t = 173.143$ GeV, which is in
agreement with the experimental value for top quark mass.

Rewriting the mathematical relation $(\ref{mtmwmz})$, we can
obtain the empirical relation given by \be \frac{m_t - (M_W +
M_Z)}{m_t}= 0.0023, \la{plus} \ee which has been a motive of
interest in references \cite{macgregor,cameron}.

In the approach of particle mass generation that we have
introduced on this paper, the top quark has acquired its mass from
the interactions with physical vacuum and the electroweak gauge
bosons have acquired their masses from the charge fluctuations of
physical vacuum. The common origin from vacuum for these particle
masses on this approach allow us to give a theoretical explanation
to the empirical mass relation given by $(\ref{plus})$. Some other
works in the literature \cite{macgregor,cameron} have also
suggested that the relation $(\ref{plus})$ is perhaps more than a
mere coincidence.


\section{Conclusions}\label{sec:05}

We have presented an approach for particle mass generation in
which we have extracted some generic features of Higgs mechanism
that do not depend on its interpretation in terms of a Higgs
field. The physical vacuum has been assumed to be a medium at zero
temperature constituted by fermions and antifermions interacting
among themselves by means of gauge bosons. The fundamental
approach describing the dynamics of this physical vacuum is the
SMWHS. We have assumed that each fermion flavor in the physical
vacuum is associated with a chemical potential $\mu_{f}$ in such
manner that there is an excess of antifermions over fermions. This
fact implies that the vacuum is thought to be a virtual medium
having a net antimatter finite density.

Fermion masses are calculated starting from the fermion
self-energy which represents fundamental interactions of a fermion
with physical vacuum. The gauge boson masses are calculated from
charge fluctuations of physical vacuum which are described by the
vacuum polarization tensor. We have used the following general
procedure to calculate these particle masses: Initially we have
written one-loop self-energies and polarization tensors at finite
temperature and density, next we have calculated dispersion
relations obtaining the poles of fermion and gauge boson
propagators, from here we have obtained the fermion and gauge
boson effective masses at finite temperature and density, finally
we have identified these particle effective masses at zero
temperature as physical particle masses. This identification can
be performed because in our approach the medium at finite density
and zero temperature represents the physical vacuum.

Using this approach for particle mass generation, we have obtained
masses of electroweak gauge bosons in agreement with their
experimental values. A further result is that left-handed
neutrinos are massive because they have a weak charge.
Additionally this approach has given an explanation to an existing
empirical relation between the top quark mass and the electroweak
gauge boson masses.

\section*{Acknowledgments}

We thank Vicerrectoria de Investigaciones of Universidad Nacional
de Colombia by the financial support received through the research
grant "Teor\'ia de Campos Cu\'anticos aplicada a sistemas de la
F\'isica de Part\'iculas, de la F\'isica de la Materia Condensada
y a la descripci\'on de propiedades del grafeno". C. Quimbay
thanks Alfonso Rueda, Carlos Avila, Miguel Angel Vazquez-Mozo,
Rafael Hurtado and Antonio S\'anchez for stimulating discussions.


\end{document}